\documentclass[fleqn,10pt]{wlscirep}
\usepackage[utf8]{inputenc}
\usepackage[T1]{fontenc}
\usepackage[normalem]{ulem}
\usepackage{amsmath}
\title{Instrumentation Prospects for Rocky Exoplanet Atmospheres Studies with High Resolution Spectroscopy}

\newcommand{\edited}[1]{#1}

\author[1,*]{Surangkhana Rukdee}
\affil[1]{Max Planck Institute for Extraterrestrial     Physics, Giessenbachstrasse, 85748 Garching, Germany}
\affil[*]{suri@mpe.mpg.de}

\begin{abstract}
Studying the atmospheres of exoplanets is one of the most promising ways to learn about distant worlds beyond our solar system. The composition of an exoplanet's atmosphere can provide critical insights into its geology and potential habitability. For instance, the presence of certain molecules such as water vapor, oxygen, or methane have been proposed to indicate the possibility of life. From an observation point of view, over the past fifteen years, significant progress has been made in characterizing exoplanetary atmospheres. This work reviews recent developments in ground-based high-resolution spectroscopic instruments that make it possible to analyze distant atmospheres in great detail. High-resolution transmission spectroscopy, one of the most effective methods used, has examined the atmospheres of Jupiter-like and is pushing towards the smaller, sub-Neptunian exoplanets. Numerous molecules have been detected using this technique, including $\mathrm{CO, H_2O, TiO, HCN, CH_4, NH_3, C_2H_2, OH}$. We explore the intriguing possibilities that lie ahead for future ground-based instrumentation, particularly in the context of detecting biologically relevant molecules within Earth-analog exoplanetary atmospheres including molecular oxygen ($\mathrm{O_2}$).
\edited{With detailed exposure time calculations for detecting $\mathrm{O_2}$ we find that at the same exposure time spectral resolution of 300,000 reaches higher significance compared to 100,000. The exposure time and therefore the needed number of transits is reduced by a factor of 4 in challenging haze and cloud scenarios.}
\end{abstract}
\begin{document}

\flushbottom
\maketitle
\thispagestyle{empty}

\section{Introduction}

Instrumentation has played a pivotal role in exoplanetary research over the last thirty years. Before 1995, our understanding of exoplanets primarily relied on philosophical and theoretical considerations \cite{Pepe2014}. Technical developments opened the possibility to detect and characterize exoplanets. Now, one of the most minute features in the cosmos, the thin atmospheres wrapped around distant exoplanets are being measured both from ground-based and space-based instruments. Observing the atmospheres around rocky planets, also known as terrestrial exoplanets, poses a significant technical challenge in the next decades.

\subsection{State of the art in instrumentation for exoplanet atmosphere studies}
\begin{figure}[h]
\centering
\includegraphics[width=0.7\textwidth]{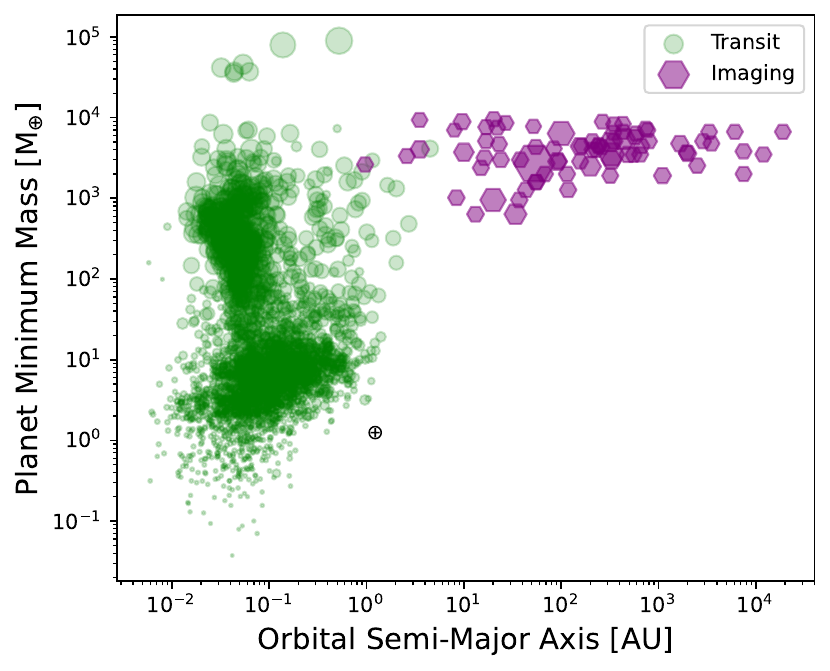}
\caption{\edited{The confirmed exoplanets discovered with high-contrast imaging (purple) are compared to those found with transits (green) as of September 2024 from the NASA Exoplanet Archive. The marker size corresponds to the relative radius.}}
\label{fig:Fig1}
\end{figure}

Observing exoplanet atmospheres presents several difficult challenges \cite{Birkby2018}. Directly seeing the emitted spectrum of exoplanets or their atmospheres is challenging due to the glare of their host stars. The star outshines the faint exoplanets with the contrast ranging from $\mathrm{10^{-3}}$ to $\mathrm{10^{-10}}$ for hot Jupiters to Earth-Sun analogs. Contrast, the brightness ratio between an astronomical source such as a planet and the star it orbits, is key for direct imaging. The ratio $\frac{F_{source}}{F_{Star}}$ is small indicating the source (planet) is much fainter than the star \cite{Follette_2023}.  Moreover, exoplanets are often located at orbital distances of sub-arcsecond from their host stars, making a spatial separation of the light, for example with coronagraphs, difficult. Direct imaging is most sensitive to planets orbiting at distances greater than 5 AU from their host stars\cite{Fischer2014}. In Figure \ref{fig:Fig1}, current directly imaged planets are shown as red squares, typically of planets of thousand times the mass of Earth. This technique enables the direct recording of photons emitted by these planets, facilitating their spectroscopic or photometric characterization. This has led to the discovery of Jupiter analogs such as 51 Eri b \cite{Macintosh_2015} and AF Lep b \cite{Franson_2023}, which are inaccessible to transit spectroscopy because their orbits are not passing in front of the host star from our point of view. The High Contrast Imaging (HCI) reflected light observation is often aided by high-contrast systems \cite{Guyon2018}, which is a combination of Extreme Adaptive Optic (ExAO) systems, coronagraphs, wavefront sensing, differential imaging techniques \cite{Follette_2023} and integral field spectrographs. The ExAO \cite{Fusco2006} helps to remove most of the optical aberrations induced by atmospheric turbulence and perform fine wavefront correction such as GPI \cite{graham2007groundbased,macintosh2014first}, SCExAO \cite{Currie_Guyon_Martinache_Clergeon_McElwain_Thalmann_Jovanovic_Singh_Kudo_2013}, and SPHERE \cite{Beuzit_2019}. The internal coronagraph applies design masks and optical components inside the telescope to induce starlight destructive interference at the expected location of a planet in the image with Lyot mask \cite{Lyot1939} or Pupil Apodization \cite{Kasdin_2003, Yang2004, Guyon_2003, Martinache2004}. The sensitivity of high-contrast imaging can be limited by speckles making diffracted starlight at the position of the planet. Speckles originate from atmospheric turbulence and optical aberrations within the instruments \cite{Ruffio2019}.  Including an integral field spectrograph helps remove speckles and acquire spectra, such as P1640+Palm3000 \cite{Dekany_2013}. The Subaru Coronagraphic Extreme Adaptive Optics (SCExAO) has achieved high-quality adaptive optics corrections with a Strehl ratio of $\sim90\%$ Strehl ratio at 1.6 $\mu$m under favorable conditions, attaining planet-to-star contrasts of  ($\sim10^{-6}$ at 0.5”) \cite{Currie2019}, comparable to those achieved by GPI \cite{Macintosh_2015} and SPHERE \cite{Chauvin_2017}. SCExAO has also reached extreme AO-like contrasts for stars as faint as $\mathrm{12^{th}}$ magnitude in the optical.

Molecule detection can be enhanced by high-resolution spectroscopy. Two decades ago, high dispersion spectroscopy (HDS) was suggested \cite{Sparks_2002, Riaud2007} as a way to boost detection capabilities by relying on the numerous spectral features expected in the spectra of exoplanets. Utilizing a high-resolution spectrograph can remove speckles by applying a high-pass filter while keeping the molecular signatures of the planet spectrum nearly intact \cite{Ruffio2019}. One of the most recent developments in high-contrast imaging for studying exoplanet atmospheres includes the incorporation of a high-resolution spectrograph in a scheme known as high-dispersion coronagraphy (HDC) to achieve a contrast of up to $10^{-7}$ \cite{Snellen2015}. HDC was first coined with an experiment\cite{Wang2017}, which demonstrated that it could achieve contrasts exceeding $10^{-7}$, depending on the target and whether observations are conducted from space or the ground. Projects employing HDC include HiRISE at VLT (combining SPHERE and CRIRES) \cite{Vigan2024}, Subaru/REACH \cite{Kotani2020_REACH}, and the Keck Planet Imager and Characterizer (KPIC) \cite{Delorme2021_KPIC} at Keck. Recent results show that they are suitable for giant planets \cite{Vigan2024, Finnerty2023CharacterizingKPIC}. 

Looking into the future, the combination of SPHERE and ESPRESSO \cite{Lovis_2017, Chazelas2022} is expected to observe $\mathrm{O_2}$ around Proxima b atmosphere within 60 nights at the Very Large Telescope (VLT) assuming a total system transmission budget of $3-4\%$. One challenge in observing Proxima b is the star-planet separation of 0.03", although this subarcsecond measurement can be achieved with an instrument similar to GRAVITY\cite{Eisenhauer2011},  the H-band instrument offers an efficiency of 1-2\% \cite{pigravity+}. The current limitation of GRAVITY will be overcome with GRAVITY+, which will enhance the detection capability of exoplanets in the J and K bands. Looking into the future, a previous study \cite{Wagner2021} cautions that with a 39-meter Extremely Large Telescope (ELT), in the case of studying the $\alpha$ Centauri AB system with high-contrast imaging, a separation of $\sim$1" would correspond to $\sim\,17.5 \lambda/D$ and would therefore likely be close to background-limited. The first-generation ELT spectrograph HARMONI\cite{Thatte2021} is expected to reach a contrast of a few $10^{-7}$ at 50 mas for bright targets in a photon noise regime with molecular mapping \cite{Bidot2024} in the range of 1.45-2.45 $\mu$m at R=18,000. With this resolution, a recent simulation for Proxima b \cite{Vaughan2024} shows that the measured spectrum is expected to be dominated by tellurics and the stellar continuum. Secondly, the first-generation ELT spectrograph METIS\cite{Brandl2021} combines high-resolution spectroscopy and high-contrast imaging. It will offer a unique capacity to characterize exoplanet atmospheres in the L and M bands, 3-5 $\mu$m range with a spectral resolution of with R=100,000 in integral field unit (IFU) mode. Further into the future, the second generation high-resolution spectrograph ANDES\cite{Marconi2022}, the 2nd generation instrument of the E-ELT, will unlock observations of terrestrial exoplanet atmospheres with high contrast and high-resolution techniques \cite{palle2023groundbreaking}.

Today, the most common method used to investigate exoplanet atmospheres involves transit spectroscopy. \edited{Transmission spectroscopy is most effective for close-in planets, as the probability and frequency of transit events decrease significantly with orbital distance \cite{Stevens_Gaudi2013}. In contrast, observations of reflected light spectra allow atmospheric characterization of planets at larger orbital separations, offering complementary insights to those obtained through transmission spectroscopy \cite{Saxena_2021}.} A challenge here is the alignment of the system so that it passes in front or behind its host star along our line of sight, known as a transit or secondary eclipse. Furthermore, the planet orbital period constrains the observing cadence by which more transit spectra can be obtained. Spectra acquired during transits can give clues on the chemical composition of exoplanet atmospheres \cite{Seager2000, Charbonneau2002}. As a planet crosses the observer's line of sight to the host star, atmospheric molecules produce distinctive spectral features through absorption lines. With the transmission spectroscopy technique, the most crucial factors are the duration and the planetary radius or atmosphere height relative to the host star radius, which sets the brightness decrease. Since timing is critical for transit studies, maximizing signal acquisition within a single transit is crucial.

High-resolution (R = 25,000 - 100,000)\cite{Birkby2018} observations improve reliability by better disentangling overlapping telluric features and concentrating spectral peaks on a few resolution elements while spreading background noise. Ground-based transit spectroscopy involves atmospheric absorption features from Earth's atmosphere (telluric features), posing the challenge of distinguishing them from extraterrestrial features. Higher resolution capabilities, ideally R$\sim$100,000\cite{Pepe2014}, are crucial for secure and efficient identification of complex spectral features, especially in separating overlapping absorption features from different molecules. \edited{Each detected line using the high-resolution technique enhances the overall planet signal-to-noise ratio by number of lines \cite{Birkby2018} within the spectral features or band (Equation \ref{eq:eq1}). The planet signal to noise ratio (SNR) is described in previous study \cite{BoldtChristmas2024} as:}
%\noindent
\edited{
\begin{equation}
SNR_p = \frac{S_p}{S_*}\,\times\,SNR_*\,\times\,\sqrt{N_{lines}}
\label{eq:eq1}
\end{equation}}
%\noindent
\edited{where $SNR_p$ is the signal-to-noise ratio of the planet, where $S_p$ represents the planet's signal, $S_*$ denotes the stellar signal, and $N_{lines}$ refers to the number of lines within the spectral feature.}

Increasing spectral resolution from R=100,000 to R=300,000 roughly doubles average line depth \cite{Currie2023} and could potentially reduce the number of transits needed for a 3$\sigma$ detection of $\mathrm{O_2}$ around terrestrial exoplanets by up to 35\% \cite{Lopez-Morales2019, Ben_Ami_2018, Currie2023}. This technique was proposed to be conducted with G-CLEF\cite{Szentgyorgyi2016} spectrograph at GMT with the assistance of the resolution booster \cite{Ben_Ami_2018, Rukdee_2020}. With both HCI and transmission spectroscopy techniques mentioned above, high spectral resolutions enhance detection capabilities for studying exoplanets and their atmospheres. 

\subsection{High-resolution atmosphere observations}
In the last decade, the High-Resolution Spectroscopy (HRS) technique has empowered astronomers to detect minute features on the surface of exoplanets. These include determining the velocity of winds\cite{Brogi2016} traversing the terminator regions of exoplanets, investigating phenomena such as thermal inversion layers and escape processes\cite{Snellen2010, Nugroho2017, YanHenning2018}, and unveiling critical information about the orbital inclinations and rotations of various exoplanets \cite{Brogi2012, Snellen2014}. Moreover, high-resolution spectroscopy has enabled the identification of molecular and atomic species within the atmospheres of diverse exoplanets \cite{Giacobbe2021, Merritt2021, BelloArufe2022} and detailed measurements of isotopologues \cite{Mollire_2019}. Precise measurements of chemical abundance can pinpoint where within the protoplanetary disk, the planets were formed \cite{Line2021, Konopacky_2013}. 

The ground-based study of exoplanet atmospheres started with the groundbreaking detection of sodium (Na) around the atmosphere of HD209458b \cite{Charbonneau2002}. In 2010, another significant milestone was reached when carbon monoxide (CO) was identified in the same planetary atmosphere \cite{Snellen2010}, marking the start of a popular observational technique known as High-Resolution Cross-Correlation Spectroscopy (HRCCS). Here, the removal of starlight and telluric features is achieved by HRCCS, which correlates the observed sequence of spectra with theoretical templates of exoplanet atmospheres. This takes advantage of different Doppler shifts of the different spectral components. HRCCS helps to disentangle the exoplanet's spectrum from the overwhelming glare of its host star, enabling the study of its physical, chemical, and biological processes. The presence of specific molecules in exoplanet atmospheres may indicate potential biological processes occurring on the planet's surface\cite{Meadows2018}. This technique can detect both transiting and non-transiting planets and provides valuable information about the planet's composition, structure, clouds, and dynamics. Recent studies targeted giant planets, led to the detection of various molecules including CO \cite{Brogi2012, Brogi2013, Brogi2014, Cabot2019, deKok2013, Rodler2013, Snellen2014}, $\mathrm{H_2O}$ \cite{Alonso-Floriano2019, Birkby2013, Brogi2014, Brogi2016, Cabot2019, Guilluy2019, Hawker2018, Lockwood2014}, $\mathrm{CH_4}$ \cite{Guilluy2019}, HCN, \cite{Cabot2019, Hawker2018}, TiO \cite{Hoeijmakers_2015,Nugroho2017}. It is also employed to identify biomarkers and map features in the atmospheres of exoplanets \cite{Birkby2018}. Six simultaneous species  \cite{Giacobbe2021} abundances, including $\mathrm{NH_3}$, $\mathrm{C_2H_2}$, and OH detection was previously reported but recently conflicted with JWST observation \cite{Xue2024} where the previous study \cite{Giacobbe2021} did not include a retrieval to formally constrain molecular abundances due to the challenges associated with analyzing high-resolution spectroscopy (HRS) data.

Clouds and haze can disrupt the spectral characteristics, resulting in a flat transmission spectrum \cite{Kreidberg2014, Hood2020}. Clouds, formed from particle condensation, exhibit much higher opacity in both absorption and scattering, compared to atmospheric gases. This diminishes the visibility of spectral characteristics, restricting telescopes' access to only the atmospheric layers above clouds \cite{Helling_2019}. Hazes are particles produced by photochemical reactions that affect opacity across the entire wavelength range. They also cause Rayleigh scattering in the optical spectrum. \edited{The featureless spectrum, despite being obtained with sufficient SNR, can be caused by clouds and hazes}, which may hinder the diagnosis of the molecular composition of exoplanet atmospheres. Several cooler and smaller exoplanets have shown spectra interpreted as caused by cloudy atmospheres, which poses challenges to constrain chemical species\cite{Helling2023}. HRS is a technique to mitigate the effects of clouds and hazes. It leverages its sensitivity to spectral line cores to probe regions situated at higher altitudes above the clouds \cite{Gandhi2020, Xuan_2022}.

For smaller planets such as mini-Neptunes and super-earths, so far only upper limits of certainty species in atmospheres\cite{keles2022} could be established \cite{Crossfield2011, Esteves2017}. The measurement is challenging because a small planet's atmosphere absorbs only less than a 1\% fraction of the star's light. When observing exoplanets, ground-based telescopes face several challenges: seasonal observability constraints, variations in weather conditions, and the day-to-night cycle \cite{Currie2023}. However, despite these limitations, ground-based telescopes have the advantage of being able to detect certain molecules, such as $\mathrm{O_2}$, in the atmospheres of terrestrial exoplanets with the upcoming ELTs instruments suite. \edited{This is due to the current generation of space-based spectroscopic instruments such as JWST lacking sensitivity in the $\mathrm{O_2}$ bands.}

The scientific case for studying $\mathrm{O_2}$ in the universe has been proposed for a long time. So far it has been found in molecular cloud \cite{Goldsmith2000,Liseau2012}, nebula\cite{Goldsmith2011} and quasi-stellar object\cite{Wang_2020} where the line can be disentangled from a telluric line due to its redshift, enables observation with ground-based millimeter facilities. In exoplanet atmospheres, ozone ($\mathrm{O_3}$) was suggested to be a proxy of oxygen ($\mathrm{O_2}$)\cite{Leger1993,DesMarais2002,Segura2003, Meadows2018}. However, the recent theoretical work \cite{Kozakis2022} emphasizes that it is extremely difficult to infer $\mathrm{O_2}$ levels from an $\mathrm{O_3}$ because the condition depends strongly on the stellar type and the incident UV flux. Utilizing a spectrograph with a resolution of R=100,000 on an Extremely Large Telescope (ELT), the endeavor to detect $\mathrm{O_2}$ in exoplanetary atmospheres within a 20-parsec radius will take approximately 60 to 100 years, after relative velocities, planet occurrence rates, and the feasibility of target observations are taken into account \cite{Hardegree-Ullman2023}. In an optimistic scenario, with the possibility of combining signals from multiple ELTs, the investigation for TRAPPIST-1d could potentially be accomplished in approximately 16 to 25 years with traditional high-resolution spectrograph on three ELTs. It is previously suggested that the resolution R=300,000-500,000 instrument\cite{Lopez-Morales2019} will be needed for the next generation ELTs to expedite the observation process. Furthermore, a concept for the next NASA flagship mission\cite{Clery2023} recommended by the National Academies’ Pathways to Discovery in Astronomy and Astrophysics for the 2020s, Habitable Worlds Observatory (HWO), also plans to study $\mathrm{O_2}$ in terrestrial exoplanets launch in the 2040s. 

To summarize, both transit spectroscopy and reflected light direct imaging techniques strongly benefit from efficient high-resolution spectrographs. This is particularly important if clouds and hazes are as abundant in atmospheres as hinted by recent results, as it also helps boost the signal within the resolved spectral lines. This work delves into the present-day high-resolution instruments employed in the study of exoplanets' atmospheres and outlines a potential solution to overcome the resolution limit for the development of future instruments designed for Earth-analog observation with High-Resolution Spectroscopy (HRS). The techniques for establishing HRS are detailed in Section \ref{methods}. Section \ref{results} presents the outcomes of the prototype for the future interferometer-based instrument, with the discussion in Section \ref{discussion}.

\section{Methods}\label{methods}
In this section, we collate state-of-the-art methods to technically achieve HRS. These are then empirically compared in the next section. HRS helps resolve molecular features into individual lines, allowing for a more detailed analysis of the exoplanet's atmosphere. To achieve the utmost precision when utilizing the Doppler Spectrograph for observing slow-rotating, solar-type stars, it is recommended to utilize a spectral resolution of no less than $\mathrm{R = \lambda/\Delta\lambda = 100,000}$ \cite{Pepe2014, NAP25187}. This level of resolution and sampling frequency not only amplifies the signal-to-noise ratio for each spectral line but also mitigates potential instrumental errors.  In making this selection, factors such as telescope mirror size, simultaneous spectral coverage, resolving power, and overall throughput must all be considered carefully. These high-resolution spectrographs can be broadly categorized into two main groups: those reliant on diffraction grating (such as Echelle and Immersion Grating spectrographs) and those based on interferometry, employing either Michelson Interferometers or Fabry-Perot Interferometers.

%non-relativistic Doppler formula R = lambda/delta lambda = c/v where v is the minimum velocity shift. At low resolution R=500, delta V=600 km/s. This generally insufficient for most astrophysical sources, R=5,000, delta v = 60 km/s, ok for broadest spectral lines.  R >= 50,000, delta v = 6 km/s, ok for most spectral lines and exoplanet orbital motion.

\begin{table}[ht]
\centering
\begin{tabular}{l|l|c|l}
\hline
Instrument & Telescope \& size & Range ($\mu$m) & Spectral Resolution  \\
\hline
NIRSPEC\cite{Mclean1998} & KECK II 10 m & 0.95 - 5.50  & 35,000\cite{Delorme2021_KPIC}   \\
UVES\cite{Dekker2000} & VLT 8 m & 0.30 - 1.10  & 30,000 - 110,000  \\
HDS\cite{Noguchi2002} & Subaru 8 m & 0.95 - 5.50  & 160,000   \\
IRD\cite{Kotani2018} & Subaru 8 m & 0.95 - 1.75  & 70,000   \\
SPIRou\cite{Moutou2015} & CFHT 3.58 m & 0.98 - 2.35  & 70,000   \\

CRIRES\cite{Kaeufl2004} & VLT 8 m & 0.95 - 5.20  & 100,000   \\
CRIRES+\cite{Dorn2023} & VLT 8 m & 0.92 - 5.20  & 100,000   \\

GIANO\cite{Oliva2018} & TNG 10 m & 0.95 - 2.50  & 25,000   \\
CARMENES\cite{Quirrenbach2014} & CAHA 3.5 m & 0.52 - 1.71  & 80,000-100,000   \\
IGRINS\cite{Park2014} & Gemini South 8.1 m & 1.45 - 2.45  & 40,000   \\
HARPS\cite{Mayor2003} & KECK II 10 m & 0.38 - 0.69  & 115,000   \\
HARPS-N\cite{Cosentino2012} & TNG 10 m & 0.38 - 0.69   & 115,000   \\
PARVI\cite{Cale2023} & Hale 5.1 m & 1.145 - 1.766  & 50,000-70,000   \\
EXPRES\cite{Blackman2020} & Lowell Discovery 4.3 m & 0.38 - 0.78  & 137,000   \\
NIRPS\cite{Wildi2022} & ESO 3.6 m & 0.95 - 1.8  & 100,000   \\
iLocator\cite{Crass2022,Bechter2018} & LBT 2$\times$8 m & 0.97 - 1.31  & 190,000   \\

ESPRESSO\cite{Pepe2021} & VLT 4$\times$8 m & 0.38 - 0.78   & 70,000 - 200,000   \\
PEPSI\cite{Strassmeier2015} & LBT 2$\times$8 m & 0.38 - 0.91  & 120,000 - 250,000   \\

\hline

\end{tabular}
\caption{\label{tab:instruments}Ground-based high resolution spectrographs can be used for the exoplanet atmospheres observations.}
\end{table}

Most of the recent HRS exoplanet atmosphere observations adopt instruments originally designed for precision radial velocity (PRV). These are optimized to conduct time series observation with long-term stability. Traditionally, the Echelle spectrograph is a preferable option for PRV high-resolution instruments.  Echelle spectrographs have hit a limit of resolutions of approximately R=100,000 for telescopes with apertures spanning $6.5-10$ meters. Some exceptional cases involve the use of a pupil slicer scheme e.g. PEPSI\cite{Strassmeier2015, Blackman2020, Wildi2022} or single-mode fiber feeding\cite{Kuo_Tiong2020,Crass2022,Cale2023}, which has the potential to enhance the resolution to as high as R=250,000. This constraint is predominantly rooted in the principle of etendue preservation ($\mathrm{E = A \times \Omega}$ where A is the beam cross-section area and $\Omega$ is the solid angle). The challenge becomes pronounced as larger telescope apertures, such as VLTs and ELTs, require the management of more extensive collimated beams, dispersion gratings, and, consequently, a bulkier and heavier spectrograph as a whole. As a result, labor, fabrication, and construction costs increase. This challenge is exacerbated when considering the next-generation Extremely Large Telescope (ELT) because the resolution is inversely related to telescope diameter, as expressed in the relationship $R\propto\frac{1}{D}$ \cite{Schroeder2000} \edited{the case of a diffraction-limited instrument}. However, currently, all the exoplanet atmosphere observations are based on Echelle or Immersion grating spectrograph as shown in Table \ref{tab:instruments} with the exception of single-model fiber-fed instruments such as iLocator\cite{Crass2022} and PARVI\cite{Cale2023}. 

For precision and stability, single-mode fibers (SMFs) are considered optimal for guiding light into spectrographs \cite{Crepp2014, Jovanovic2016}. The SMF spectrographs operate in the diffraction-limited regime, achieving exceptionally high spectral resolution in a compact instrument. It allows the study and mitigation of stellar variability \cite{Crass2022}. However, efficiently coupling a point spread function (PSF) into an SMF poses inherent challenges due to their intrinsic properties \cite{El_Morsy_2022}. The theoretical maximum coupling efficiency of an unobstructed circular aperture into an SMF is approximately 81\% \cite{Shaklan1988}. In practice, most ground-based telescopes feature obstructed circular apertures and spiders in the pupil, which significantly impact coupling efficiency, for example, resulting in about 60\% efficiency at telescopes like Keck and Subaru \cite{Jovanovic2017} and approximately 73\% at VLT \cite{Otten2021}. HiRISE needs a coupling efficiency exceeding 95\%\cite{Otten2021} to maximize feasibility to spectrally characterize exoplanets. Achieving this objective requires precise centering of the PSF with the core of the fiber within 0.1 $\lambda/D$ \cite{El_Morsy_2022}, which is extremely challenging. \edited{A coupling efficiency as low as 59\% of the maximum is acceptable for HiRise, corresponding to a misalignment of 0.2 $\lambda/D$ between the PSF and the fiber. In this case, characterizing the most challenging targets would be unfeasible without the required additional telescope time.} KPIC reported achieving the placement of the planet's PSF on the science fiber with a precision of less than 0.2 $\lambda/D$ ($<$ 10 mas)\cite{Delorme2021_KPIC} in the K band. 

Fourier Transform Spectrometers (FTS) are based on the principle of the Michelson interferometer. They derive spectral information via an interference-based process and mathematical algorithms. An illumination source is split into two beams via a beam splitter, creating a path difference. The beams recombine, forming an interferogram on a detector. Fourier Transformation converts this time-domain interferogram into a spectrum, revealing spectral properties\cite{Davis2001}. FTS instruments are complex, sensitive to vibrations, and substantial in size \cite{Hall1979}. Achieving higher resolution often necessitates elongated optical paths, which can, in turn, impose limitations on the range of accessible wavelengths. The spectral resolution of FTS is determined from the Optical Path Difference (OPD). Primarily owing to their subpar system efficiency of broadband spectroscopy, the use of traditional FTS in stellar astronomy has been limited \cite{Behr2009}. The overall on-sky efficiency of the dispersed Fourier transform spectrograph (dFTS), at a spectral resolution of R = 50,000, reaches a level of efficiency (4-10\%)\cite{Behr2009} similar to that of Echelle spectrographs. Conversely, FTS is highly effective for solar observations, achieving ultra-high spectral resolutions of R = 700,000 to 1,000,000, such as FTS instruments at Kitt Peak \cite{Wallace_2011} and an FTS from Institut für Astrophysik, Göttingen (IAG) \cite{Reiners2016}. 

The Virtually Imaged Phased Array (VIPA)\cite{VIPA} functions as an angular dispersive device similar to a prism and diffraction grating. It separates light into its spectral components and operates almost independently of polarization. Unlike prisms or conventional diffraction gratings, the VIPA exhibits significantly higher angular dispersion while possessing a smaller free spectral range. It also offers structural simplicity and compact dimensions within a reasonable cost range. VIPA is well-suited to serve as the primary dispersion element in ultra-high resolution spectrometers for both ground \cite{Zhu_2020} and space instruments \cite{Bourdarot2017}. Increased spectral resolution may result in decreased wavelength coverage. VIPA spectrographs are well-suited for ultra-high resolution applications but are constrained to a limited spectral range of approximately 200 nm. This spectral limitation may pose challenges when simultaneous observation across a broad wavelength range is required. VIPAs, akin to Fabry-Perot Interferometers (FPI), exhibit sensitivity to input light's spectral characteristics, introducing potential complexities in system alignment. Furthermore, VIPAs can generate intricate fringe patterns in their spectra, demanding specialized analysis methods. Nevertheless, the compact size of VIPA allows for versatile implementation. VIPA technology offers versatility in its application, accommodating both single-mode and multi-mode fiber systems \cite{Zhu2023}, with a potential throughput of up to 40\% \cite{Carlotti2022}. \edited{Currently, VIPA spectrographs\cite{Zhu2023,Carlotti2022} employ diffraction gratings as cross-dispersers. Further improvements in throughput can be achieved by using VPH cross-dispersion.}

Current generation ground-based and space-based instrumentation are not yet capable of studying $\mathrm{O_2}$ around Earth-analogs. Alternative approaches and novel instrumentation concepts have been introduced to study $\mathrm{O_2}$ around Earth-analogs with ELT.  Firstly, the Fabry Perot Instrument for Oxygen Searches (FIOS) \cite{Ben_Ami_2018, Rukdee2023} is an FPI based resolution booster designed as a plug-in for the traditional (external) high-resolution spectrograph being built for ELTs to study molecular oxygen with transmission spectroscopy and cross-correlation technique. It splits the signal into multiple, each with a different (shifted) periodic high-resolution profile imposed. Analyzing the spectra of these signals with an external spectrograph gives a higher resolution. It also improves the spectral sampling frequency of the spectral profiles with the chained FPI. In the prototype, the demonstration of FIOS subunit was plugged into a VIPA-based external spectrograph. This resulted in a very compact instrument. This solution could reduce the observation time on the ELT to at least half \cite{Hardegree-Ullman2023} from the signal gain into the high-resolution feature. Secondly, a new collaboration established in 2023 proposed studying $\mathrm{O_2}$ in the atmosphere of the nearest rocky exoplanet, Proxima b, and its analogs using Ultra-fast AO Technology Determination for Exoplanet Imagers from the Ground (UNDERGROUND) \cite{Fowler2023}. This serves as a test bed for high-contrast imaging and direct spectroscopy. The instrument is designed to achieve a contrast goal of  $3\times10^{-5}$ at 10 $\lambda /D$, with a combination of ground-based extreme adaptive optics and high-resolution VIPA spectrograph\cite{Carlotti2022}.

\edited{In this work, an exposure time calculator (ETC) was developed to numerically assess the benefits of high-resolution spectroscopy for detecting molecular oxygen in exoplanet atmospheres, comparing oxygen-rich and oxygen-free models at the $\mathrm{O_2}$ A-band. The calculation method follows the IGRINS' ETC\cite{Le2015}, which has already been applied to other instruments \cite{Rukdee_TARdYS_2019, Vanzi2018}, with adjustments for high-resolution features by summing the signal from all spectral lines within the oxygen A-band. The $\mathrm{O_2}$ ETC is optimized for simulating the signal-to-noise sensitivity per resolution element at the Oxygen A-band, comparing an oxygen atmosphere model with a no-oxygen atmosphere. In addition to the cross-correlation technique, the likelihood method performs comparably well for template comparisons\cite{Hood2020}. In this simulation, the process involves comparing two models: one containing $\mathrm{O_2}$ absorption features and another without. The code calculates the difference in chi-squared values ($\mathrm{\Delta\chi^2}$) between these model spectra calculated with ESO SkyCalc \cite{Noll2012,Jones2013,Moehler2014}.} 
%
%\noindent
\edited{\begin{equation}
\Delta\chi^2 = \Sigma(\frac{data-signal_{NoO_2}}{Noise})^2-\Sigma(\frac{data-signal_{O_2}}{Noise})^2)
\label{eq:eq2}
\end{equation}}
\edited{A Monte Carlo method is used to generate simulated data by sampling from a normal distribution centered on the signal with $\mathrm{O_2}$ and the given noise level. The $\mathrm{\Delta\chi^2}$ is computed for each simulated dataset showing the difference between the chi-squared values for the no-$\mathrm{O_2}$ model and the $\mathrm{O_2}$ model as shown in Equation \ref{eq:eq2} and averaged over 1000 data realizations.}

\edited{This simplified calculation does not account for telluric contamination in ground-based observations, making it applicable for both space- and ground-based observations. Telluric effects were addressed in a previous study\cite{Currie2023}. Here the calculation is focused on half of the Oxygen band (0.766 - 0.772 $\mathrm{\mu}m$), where the O2 absorption with the airmass=1 model of ESO SkyCalc is not saturated to complete absorption. Nevertheless, our exposure estimates should generalize approximately proportionally to the full oxygen band. The calculation incorporates realistic parameters for modern spectrograph components, such as low read-noise (1 electrons) CMOS detectors and reduced dark current (0.1 electrons/sec) levels due to detector cooling. The calculation assumes an 8-meter telescope with different instrument resolutions assuming 1\% transit depth with thermal noise excluded, as the effect start from H band (1.4 $\mu$m) onward \cite{Noll2012}.}

\edited{The comparison using the exposure time calculator focuses solely on the instrument's signal for oxygen observation. In practice, spectral information from exoplanet atmospheres is further degraded by telluric lines (from Earth). Simulations have already been conducted to account for these effects, such as different levels of cloud impacts \cite{Currie2023} and haze, which can suppress up to 50\% of spectral features \cite{Hood2020} in the case of 3$\times$ planet's isolation. The previous work \cite{Hood2020} presented several haze scenarios motivated by observations of featureless spectra. Atmospheric transmissivity (shown in Figure 3 for Earth) is affected such that the upper half is cut off and set to a constant (see Figure 4 in the previous work\cite{Hood2020}). Similarly, the simulation here considers three scenarios: a clear atmosphere without haze or clouds (optimistic case), 50\% haze (realistic case), and 90\% haze (pessimistic case). The percentage indicates how the $\mathrm{O_2}$ transmissivity is clipped. Figure \ref{fig:Fig3} left panel illustrates the assumed transit depth for the three scenarios (blue, purple, and red curves).}

\section{Results} \label{results}

In this work, we compare and contrast current and upcoming dispersing techniques for high-resolution spectroscopy. This section reviews Echelle-based and interferometry-based instruments, such as Fourier Transform Spectrograph (FTS) and Fabry-Perot Interferometer (FPI). In the results section, we compare their effectiveness in achieving exceptionally high-resolution spectral profiles empirically. For this purpose, we compare telluric oxygen observations taken by an Echelle spectrograph, X-shooter, to assess the level of low-medium resolution against the ultra-high resolution from FIOS-demo and FTS spectrograph. We note here that typical Echelle spectrographs nowadays can reach R=100,000, as shown in Table\ref{tab:instruments}. In some exceptional cases, it is limited to R=200,000.

\begin{figure}[h!]
\centering
\includegraphics[width=\linewidth]{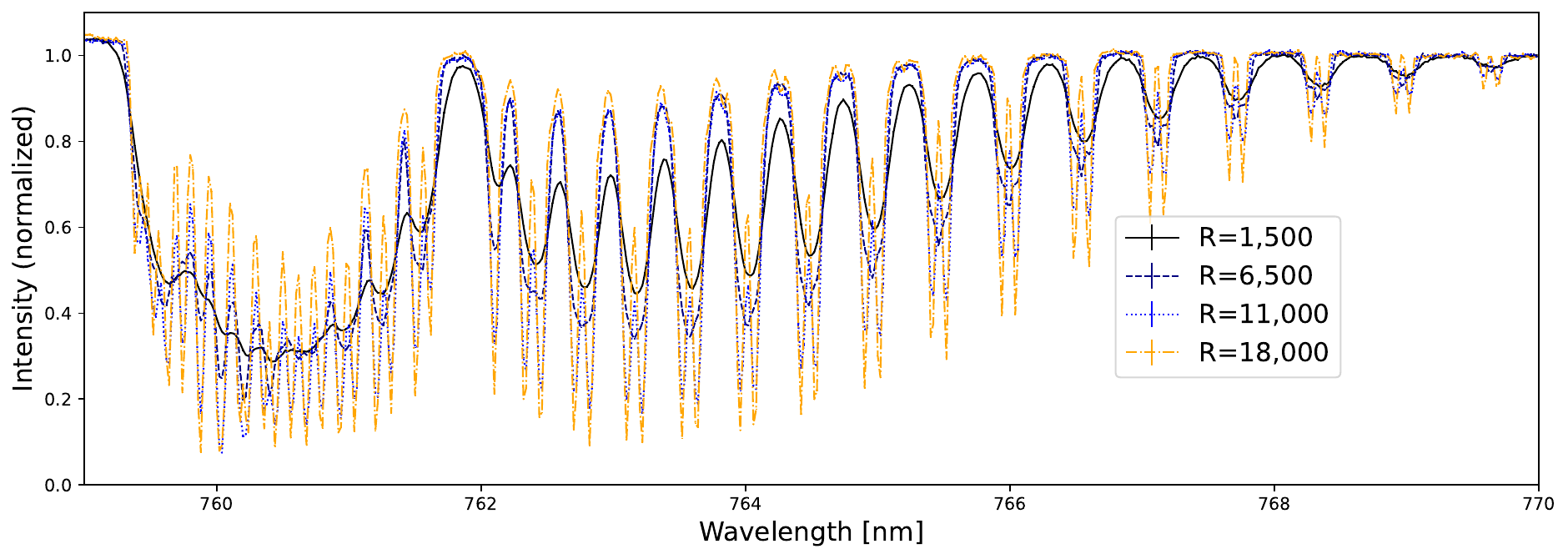}
\includegraphics[width=\linewidth]{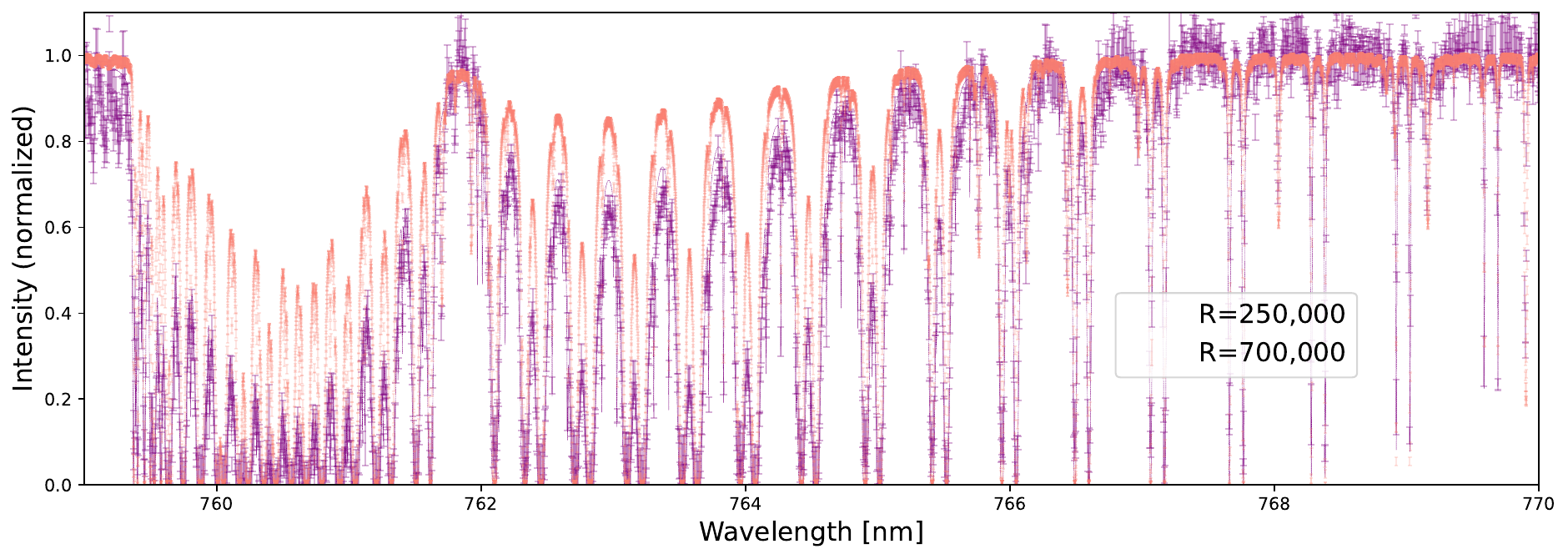}
\caption{Comparison of (telluric) molecular oxygen spectrum observed using low-medium resolution spectrograph and interferometer-based instruments. The top panel shows the observed telluric spectrum from archival X-Shooter spectra with spectral resolution ranging from R=1500 to R=18000 in different colors. The lower panel shows the spectrum from FTS\cite{Wallace_2011} (red) with R=700,000 and FIOS-demo\cite{Rukdee2023} (purple) with R=250,000 overlaid with a fitted model. Note that the two observations were taken at different altitudes, which explains the differences in the line depths of the features.}
\label{fig:Fig2}
\end{figure}

\begin{figure}[h!]
\centering

\includegraphics[width=0.6\linewidth]{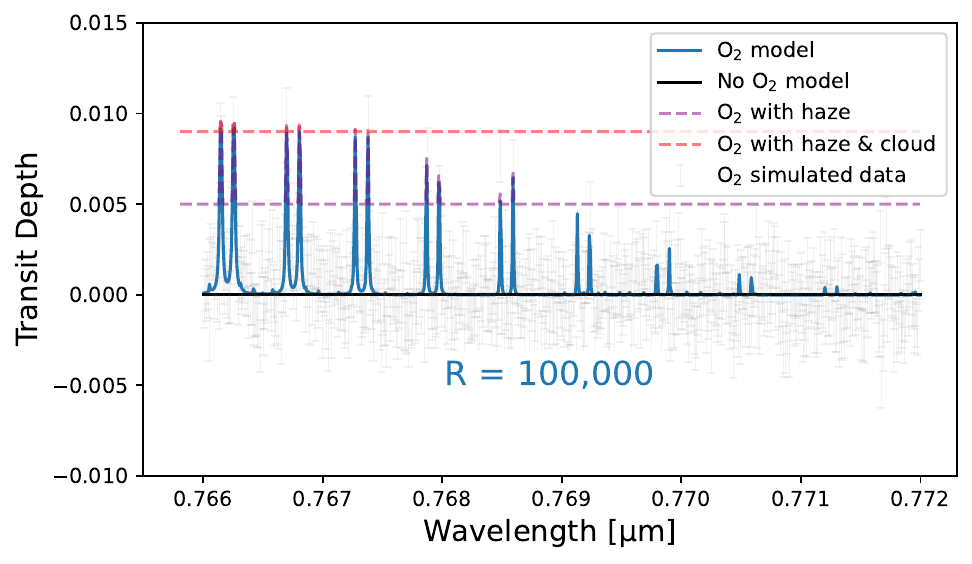}
\includegraphics[width=0.38\linewidth]{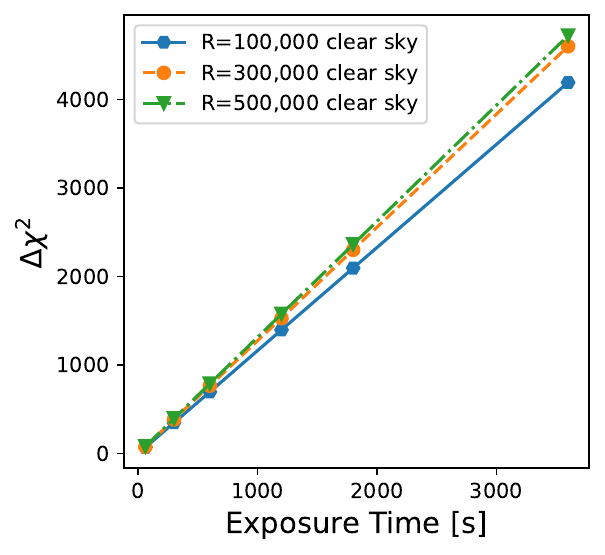}
\includegraphics[width=0.6\linewidth]{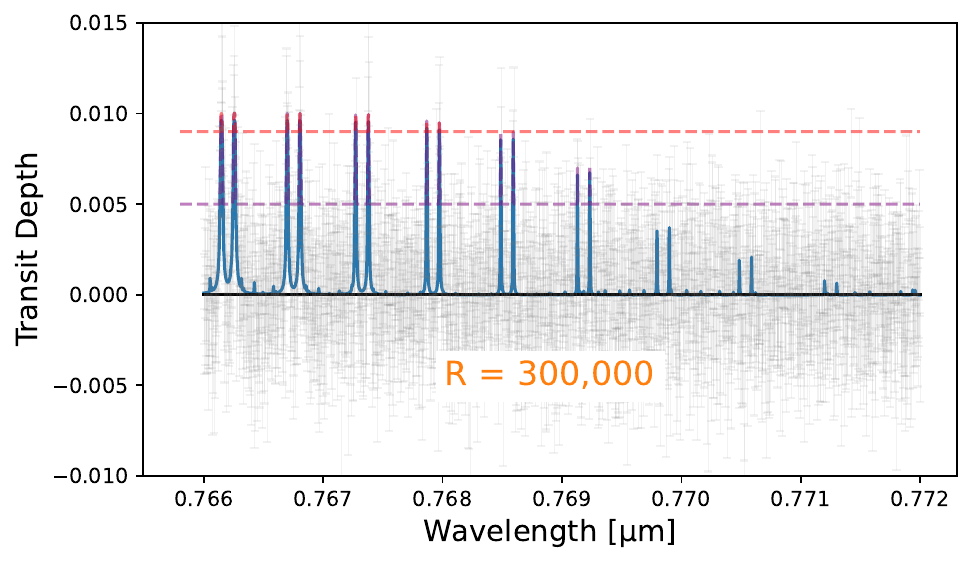}
\includegraphics[width=0.38\linewidth]{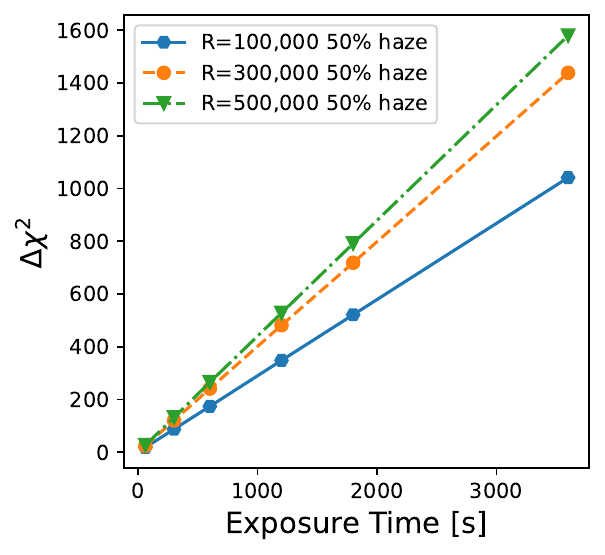}
\includegraphics[width=0.6\linewidth]{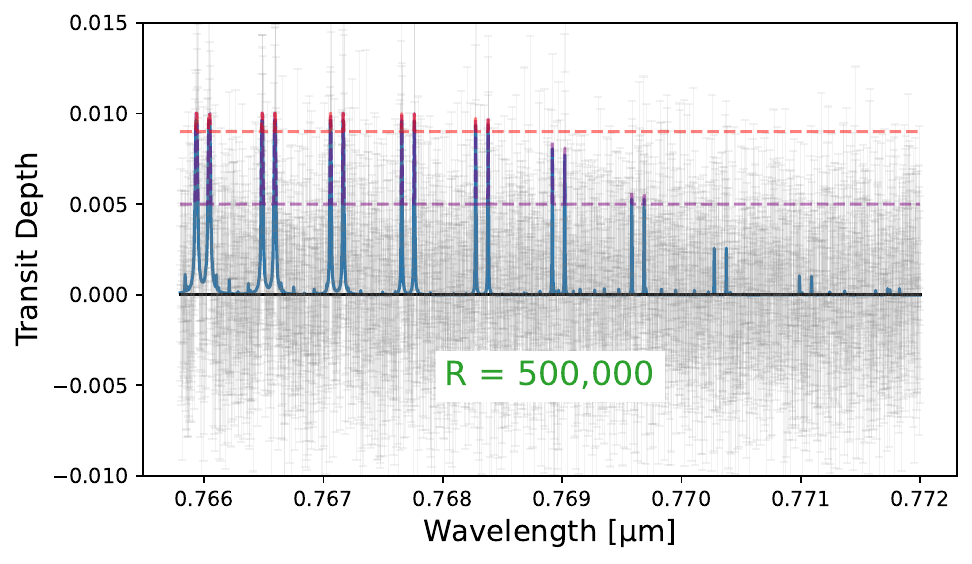}
\includegraphics[width=0.37\linewidth]{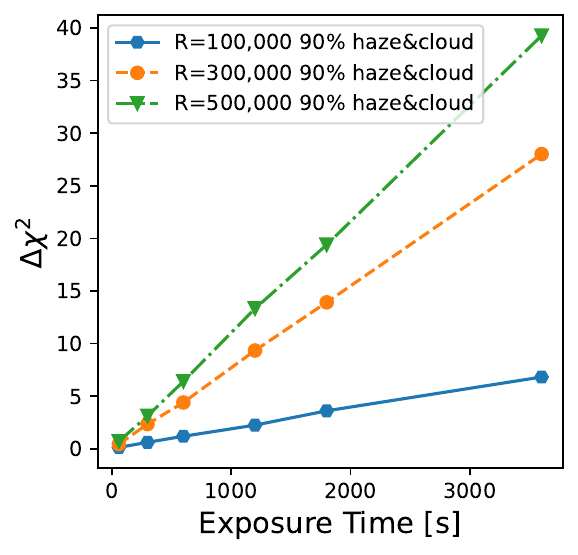}
\caption{Comparison of transit depth observed at different spectral resolutions: R=100,000 (top left panel, blue line in the right panels), R=300,000 (middle left panel, orange line in the right panels), and R=500,000 (bottom left panel, green line in the right panels), assuming 1\% of the star's light passes through the exoplanet's atmosphere. Left panels: simulated model comparison of transit spectral features for different scenarios—No-$\mathrm{O_2}$ (black solid line), clear sky $\mathrm{O_2}$ (blue solid line), simulated data for the corresponding resolution (grey data points), signal clipped by 50\% due to haze (purple dashed line), and signal clipped by 90\% due to haze and cloud deck combined (red dashed line). Right panels: model comparison confidence with the different resolutions as the exposure time increases. The three observing conditions, with clear sky (top right panel), 50\% haze (middle right panel) and haze\&cloud (bottom right panel}
\label{fig:Fig3}
\end{figure}

Figure \ref{fig:Fig2} illustrates a comparison between the observed low-medium resolution and the high-resolution spectral profiles of the oxygen A band, depicting observations of (telluric) molecular oxygen. The upper panel of Figure \ref{fig:Fig2} shows low to medium resolution telluric oxygen features. These were obtained from the ESO Science Archive Facility using X-shooter\cite{Vernet_2011} observations during February and March 2024 by the UVES team, as part of Program ID: 60.A-9022(c), OB ID:2024672, 2024624 and 2024822, at various resolutions with short exposures (12 seconds). The results indicate that higher resolution enables the observation of more detailed features within the molecular oxygen spectrum, revealing the signal more distinctly within each spectral line. The lower panel of Figure \ref{fig:Fig2} shows performance tests for future HRS instrumentation by observing the Sun through the Earth's atmosphere. These profiles demonstrate the measurement outcomes obtained using two types of interferometers: Michelson-based and FPI-based. Firstly, the FTS from the National Solar Observatory at Kitt Peak \cite{Wallace_2011} reported R=700,000 in the oxygen A-band. Secondly, the FIOS-demo\cite{Rukdee2023} showcases spectral profiles based on a chained FPI array with a spectral resolution of R=250,000. This resolution can potentially increase up to R=350,000 with the addition of each array. The throughput of each additional unit, however, decreases by 10-15\% \cite{Rukdee_2020}. One benefit of achieving this level of resolution is the increase in signal-to-noise ratio and the sampling frequency for each spectral line, which may reduce the required observing time, as predicted in \cite{Currie2023, Hardegree-Ullman2023}.

\begin{figure}[h!]
\centering
\includegraphics[width=0.6\linewidth]{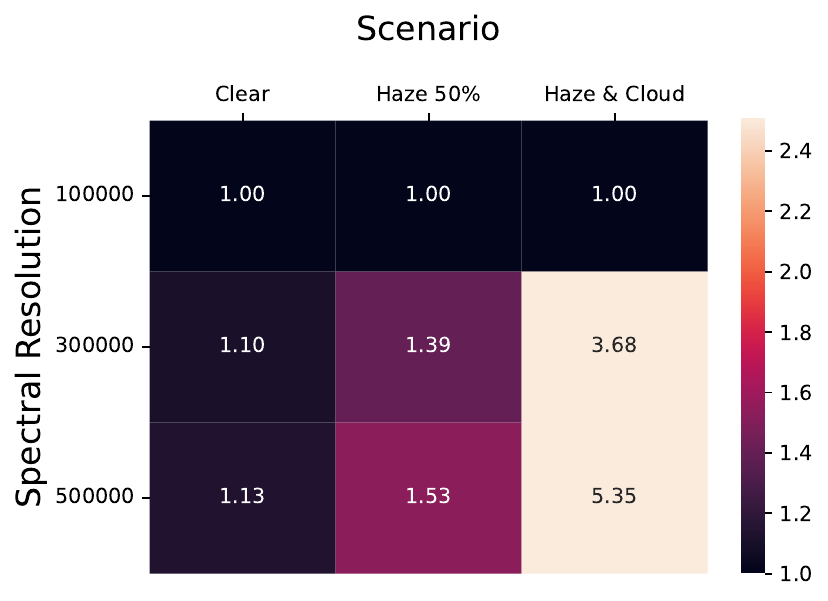}
\caption{Exposure time reduction factors. Columns indicate the considered scenarios: clear atmosphere (optimistic case), 50\% haze (realistic case), and 90\% haze (pessimistic case) at different instrument spectral resolutions (R=100,000, R=300,000, R=500,000) with a 10-minute exposure. The ratio represents the performance of ultra-high resolutions compared to R=100,000.}
\label{fig:Fig4}
\end{figure}

\edited{The results from the exposure time calculator are presented in Figure \ref{fig:Fig3}. The left panels display the non-saturated region of the oxygen A-band signal at different resolutions: R=100,000, R=300,000, and R=500,000.  These are compared with scenarios where haze reduces the signal by 50\% (50\% haze), and haze combined with a cloud deck suppresses 90\% (90\% haze and cloud) or more of the signal. The figure demonstrates that higher resolution provides more data points, thereby increasing the sampling frequency, as indicated by the increasing number of grey data points with higher resolution. For the model template matching, the right panels show the likelihood results ($\mathrm{\Delta\chi^2}$) for different cloud and haze scenarios on the planet. For clear skies, increasing the resolution from R=100,000 to R=300,000 enhances $\mathrm{\Delta\chi^2}$ by approximately 9-10\%, with a further increase to R=500,000 improving it by 11-12\%. Conversely, the exposure time can be reduced with higher resolution by these factors, because the relation between exposure time and $\mathrm{\Delta\chi^2}$ significance level is linear.}

\edited{The impact of resolution becomes more pronounced under hazy and cloudy conditions. In the realistic 50\% haze scenario where only half of the O2 signal is observable, increasing the resolution from R=100,000 to R=300,000 reduces the necessary exposure time by about 34\%, and to 50\% at R=500,000. In the most challenging scenario, where less than 10\% of the signal is observable due to haze and cloud cover, resolution increases from R=100,000 to R=300,000 can enhance 3.7 times of the signal and about 5 times at R=500,000. This is also summarized in Figure \ref{fig:Fig4} as a ratio selected for the case of 10 minute exposures. Other exposures give similar results since the linear relationship results in similar outcomes for each exposure time interval. The current calculated range focuses only on the non-saturated portion (half of the oxygen band) due to limitations in the simulated data, where the saturated part cannot be accurately compared across different resolutions. If the entire band could be observed without saturation, these ratios are expected to improve further.}

\section{Discussion}\label{discussion}
%\subsection{Near-term new instrumentation capabilities}
\begin{figure}[ht!]
\centering
\includegraphics[width=\textwidth]{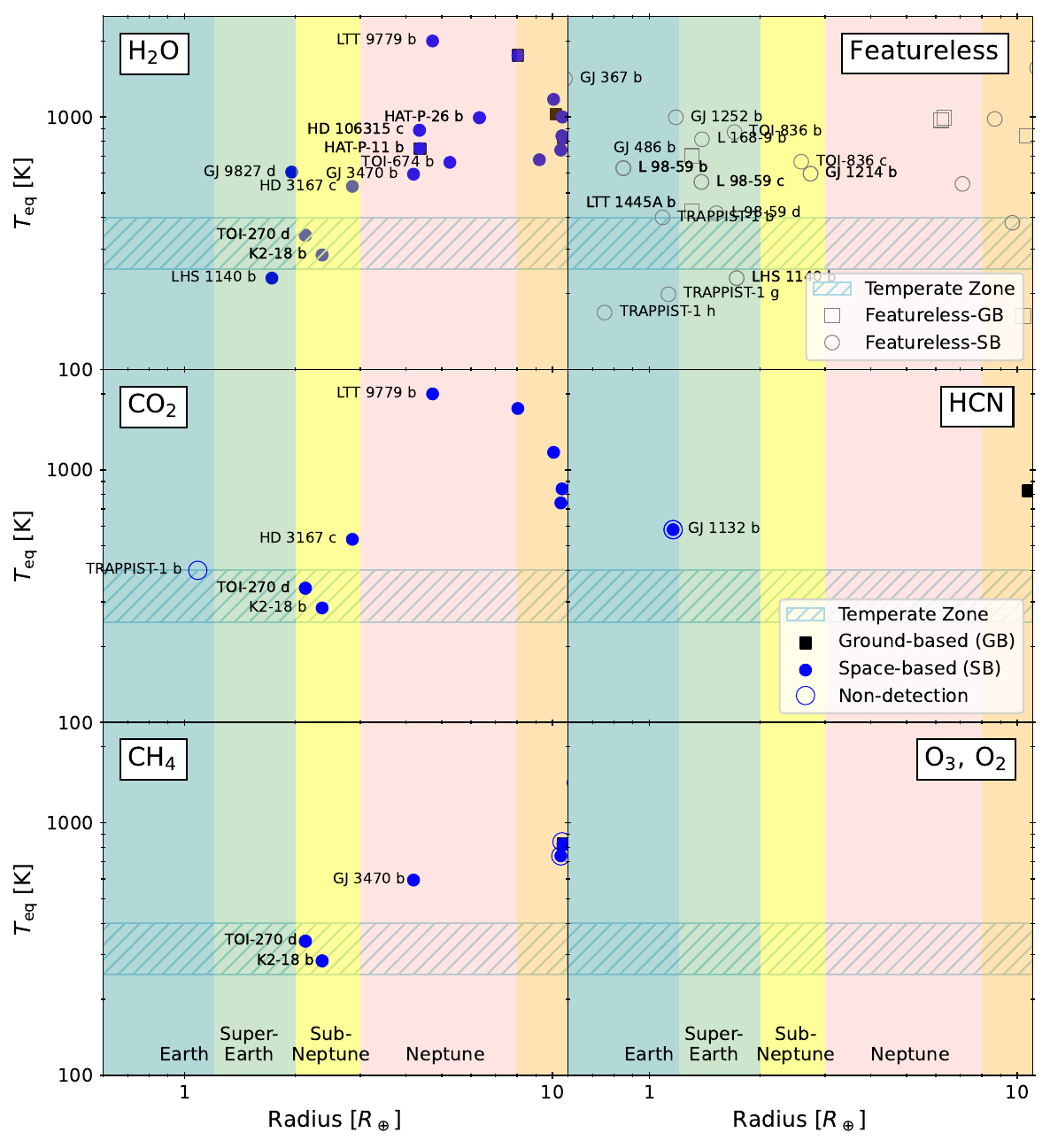}
\caption{Detection of molecules prominently involved in biological processes on Earth using current instrumentation: $\mathrm{H_2O}$\cite{Tsiaras2018,Tsiaras2019,Edwards2023,Kreidberg2022,Mikal-Evans2023, Roy2023, Edwards2021,Brande2022,Bean2023}, $\mathrm{CH_4}$\cite{Swain2021}, $\mathrm{CO_2}$\cite{Madhusudhan2023,Edwards2023, Bean2023}, HCN\cite{Tsiaras2016}, $\mathrm{O_2}$, $\mathrm{O_3}$ and featureless of Earth- to Neptune-size planets using ground-based (black square) and space-based (blue circle) instruments. Data retrieved from \texttt{ExoAtmospheres Database} as of September 2024}
\label{fig:Fig5}
\end{figure}

\subsection{Performance of current generation instruments}
Ground-based and space-based observations offer complementary insights. By combining low- and high-resolution observations, we can attain more precise and accurate insights spanning from the lower to the upper atmosphere\cite{Brogi_2017}. High-resolution ground-based observations provide a distinctive window into the thermospheres of exoplanets and exhibit sensitivity to alkali element abundances. This could be meticulously analyzed in conjunction with lower-resolution observations\cite{Pino_2018b}. While space-based instruments mitigate the telluric interference issue, their current instrumental constraints (typically R=100 up to R$\sim$3,000) restrict their ability to capture detailed spectral features. The Wide Field Camera 3 (WFC3)\cite{dressel2023} on the Hubble Space Telescope (HST) remains one of the most successful instruments for studying exoplanet atmospheres, including alkali element analysis \cite{Charbonneau2002}, particularly in the Neptune and sub-Neptune regime. Meanwhile, JWST is poised to advance the characterization of Earth and Super-Earth regimes.

Upcoming space-based instrumentation is dedicated to  exoplanetary atmospheres using low-resolution spectrographs. These include the Colorado Ultraviolet Transit Experiment (CUTE)\cite{Fleming2018}, the Atmospheric Remote-Sensing Infrared Exoplanet Large-survey mission (ARIEL)\cite{Tinetti_2016} and \edited{Twinkle \cite{Stotesbury2022}}. The CUTE CubeSat project concentrates on the near-ultraviolet region (0.25 to 0.33 $\mu$m) with a spectral resolution of 3,000. In the near future, the ARIEL mission, scheduled for a 2029 launch, focuses on a broader wavelength range spanning from 0.5 to 7.8 µm, albeit with a maximum resolution of R = 100. This mission has the ambitious goal of studying 1,000 large planets, including a select few Earth-sized ones\cite{Tinetti_2016}. \edited{The Twinkle mission will be equipped with a 0.45 m telescope and a spectrometer, providing simultaneous wavelength coverage from 0.5 to 4.5 $\mu$m.} Currently, the transit photometry monitoring space mission, TESS\cite{Ricker2014} and CHEOPS\cite{Rando_2020} offer a wealth of suitable targets for forthcoming instruments tailored to the spectral characterization of exoplanetary atmospheres.

The time is ripe for considering the initiation of a specialized effort aimed at the thorough characterization of prominent molecules which are measurable indicators of biological processes, conditions, or so-called biomarkers. 
A  theoretical study \cite{Meadows2018} proposed that the concurrent identification of $\mathrm{O_2}$, $\mathrm{O_3}$, $\mathrm{H_2O}$, $\mathrm{CO_2}$ and $\mathrm{CH_4}$ in the absence of CO might signify conditions reminiscent of those found on Earth. Remarkably, the first detection of these biomarkers around small planets using space-based instrumentation have been reported. Figure \ref{fig:Fig5} illustrates the current study of biomarkers by reported detection as listed in the \texttt{ExoAtmospheres Database}. Detections are more abundant in gas giants and at high surface temperatures, corresponding to larger stellar transit absorption signals. The dearth of detections in super-Earths and sub-Neptunes highlights the ongoing challenges. Focusing on the detection of molecules to-date in small-sized planets, $\mathrm{H_2O}$ has been detected in several Neptune-sized planets: HAT-P-26b \cite{Tsiaras2018,Athano2023}, HD106315c\cite{Kreidberg2022}, HAT-P-11b \cite{Tsiaras2018, Basilicata2024}, TOI-674b \cite{Brande2022}. The 3$\sigma$ detection of $\mathrm{H_2O}$ on LTT9779b \cite{Edwards2023} with Hubble Space Telescope's HST/WFC3, appears featureless\cite{Radica2024} with JWST Near-Infrared Imager and Slitless Spectrograph (NIRISS). It is also detected in sub-Neptune planets TOI-270d \cite{Mikal-Evans2023, Benneke2024,Holmberg2024}; K2-18b \cite{Benneke2019,Tsiaras2019}. A recent study on K2-18b \cite{Madhusudhan2023} suggests it is a Hycean planet. For super-Earth, $\mathrm{H_2O}$ is detected in GJ9827d \cite{Roy2023} and LHS1140b \cite{Edwards2021} with HST while using JWST\cite{Damino2024} shows no absorption feature although data favor an $\mathrm{N_2}$-dominated atmosphere with $\mathrm{H_2O}$ and $\mathrm{CO_2}$. The search for $\mathrm{CO_2}$ has been found in K2-18b \cite{Madhusudhan2023}, TOI-270d \cite{Benneke2024, Holmberg2024} and  LTT9779b with 2.4$\sigma$ detection from HST/WFC3 G102 (R=210 at 1000 nm) and G141 (R=130 at 1400) grisms $\mu$m but appear featureless \cite{Radica2024} due to cloud from one transit observation with JWST NIRISS (R=700) slitless spectroscopy mode. The observations on the Earth-size planet TRAPPIST-1b found little to no atmosphere and no $\mathrm{CO_2}$\cite{Greene2023} was detected with JWST-Mid-Infrared Instrument (MIRI) and featureless\cite{Lim2023} spectra with JWST NIRISS. $\mathrm{CH_4}$\cite{Benneke2024, Holmberg2024} were detected in TOI-270d with JWST-Near-Infrared Spectrograph (NIRSPEC) (R=1000-2700) and K2-18b \cite{Bezard2020,Madhusudhan2023} with both HST and JWST NIRISS. For HCN detection on 55 Cancri e\cite{Tsiaras2016} with HST but not yet seen from ground-based observation\cite{Deibert2021}. The detection on GJ1132b has been under debate\cite{May2023_GJ1132, Swain2021, Mugnai2021}. A few studies have reported the detection of CO in Neptune size planets e.g. detection in GJ3470b \cite{Beatty2024} and upper limit in GJ436b \cite{Grasser2024}.  However, this is not entirely unexpected; for instance, in Earth-like scenario, the presence of CO in the atmosphere is generally undesirable\cite{Meadows2018}. The featureless corner reported attempts to measure rocky planets with featureless spectra \cite{Barclay2023, Damiano2022, Zhou2022, Ridden_Harper2023, Bouwman2023, Zhang2024, Diamond-Lowe2023, Garcia2022, Lim2023, Moran2023}. This could be interpreted as: 1) it's cloudy and hazy, 2) there are no atmospheres around these rocky bodies due to the host star, and 3) we have reached the instrument's limit, since the rocky atmospheres are supposed to be so small compared to the rocky planet's radius. From Figure \ref{fig:Fig3}, we can also observe that the current limitations of ground-based observations rarely extend to measurements below those of Neptune size planets except for the LTT1445Ab\cite{Diamond-Lowe2023} and GJ486b\cite{Moran2023} with featureless spectra. \edited{This study demonstrates that high spectral resolution is particularly advantageous in hazy and cloudy scenarios. Future instruments should prioritize a combination of large aperture, high throughput, and high resolution to further enhance observational capabilities.}

\subsection{Future development for $\mathrm{O_2}$ detection}
Motivated by the apparent ubiquity of haze and clouds, we demonstrate the need for R\>100,000 spectroscopy with detailed exposure time simulations. In  our optimistic, realistic and pessimistic scenarios, R>300000 spectrographs give a exposure time reduction of 1.1, 2x and 4x compared to R=100000 spectrographs for the detection of $\mathrm{O_2}$.

\begin{figure}[h!]
\centering
\includegraphics[width=0.6\textwidth]{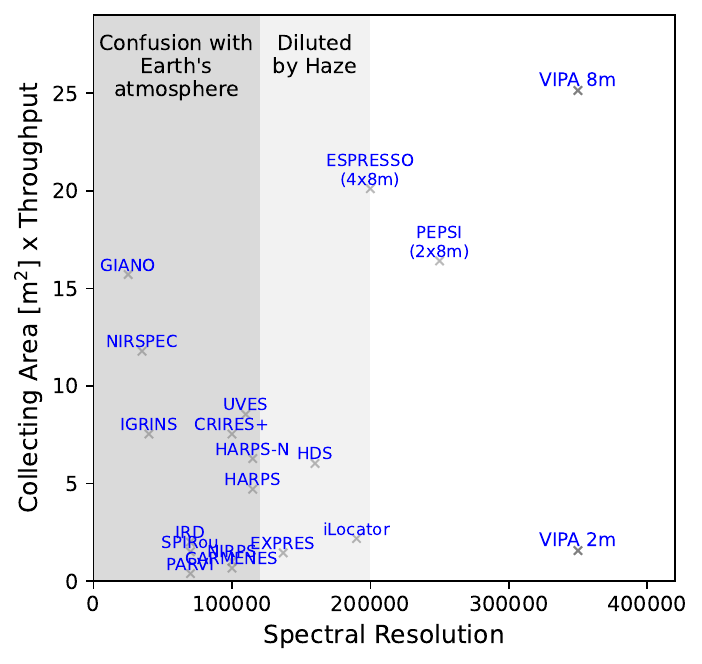}
\caption{The photon efficiency of high-resolution spectrographs, most of which are grating-based, can reach 50\% when using VIPA \cite{Carlotti2022}. The light shaded area below R=200,000 indicates where the signal may be clipped due to hazes, while the dark grey shaded area below R=150,000 suggests where the signal could be mixed with telluric features. These spectrographs, particularly those in the middle to right side of the image, open new discovery opportunities for detailed signal analysis of small planets, achieving both high resolution and photon efficiency.}
\label{fig:Fig6}
\end{figure}

\edited{With detailed and realistic exposure time calculations, we confirm and extend previous findings \cite{LM2019, Hood2020, Currie2023, Hardegree-Ullman2023} that HRS improves biosignature detection also in the Oxygen A-band. One early study \cite{LM2019} examined different oxygen bands at varying resolutions using model spectra for different stellar types suggesting that ultra-high resolution could reduce over 30\% in the number of transits required to produce a detection. A subsequent study \cite{Hood2020} simulated realistic Earth-like planet spectra, focusing on molecules other than oxygen, with resolutions up to R=100,000 using existing telescopes with possible hazy scenarios. Similarly, the third study \cite{Currie2023} modeled the ELTs' performance for transit spectroscopy at R=100,000, suggesting that increasing the resolution to R=300,000 doubles the line depth compared to R=100,000 and can also reduce the required transit time by 30\%. However, the spectral lines do not gain much additional information by further increasing the resolution to R=500,000. Finally, the most recent work \cite{Hardegree-Ullman2023} explored a survey of molecular oxygen using ELTs, comparing resolutions of R=100,000 and R=500,000 for clear sky conditions. In hazy and cloudy conditions, as shown in Figure \ref{fig:Fig3}, ultra-high resolutions of R=300,000-400,000 demonstrate significant advantages in capturing diluted signals. This calculation is applicable to ground-based high-resolution observations of small planets using both transmission spectroscopy and reflected light, as well as space-based observations to consider S/N for both cloud and haze scenarios. A key unknown affecting bio-signature detectability is the propensity of haze and clouds on exoplanets.} 

The findings from Figure \ref{fig:Fig2} demonstrate empirically the current technological range of obtaining high-resolution spectra of exoplanet atmospheres using interferometry-based instruments. The FIOS-demo employed the chained Fabry-Perot interferometer to demonstrate the booster function where the external spectrograph is VIPA-based. The simulation study \cite{Lopez-Morales2019} found that the optimal range for observing $\mathrm{O_2}$ lies between R = 300,000-400,000. The resolution beyond 500,000 would not gain much more signal into line profiles. The comparison of observational results presented here between the FIOS-demo and FTS aligns well with these theoretical predictions compared to the observed lower-resolution spectral profiles.

To obtain robust insights into the atmospheres of exoplanets as well as emission and absorption lines study, developing a high-resolution space-based mission \cite{theluvoirteam2019luvoir, gaudi2020habitable} with a large collecting area is imperative. Space-based observations of exoplanets provide a unique platform for conducting detailed and uninterrupted studies of these distant worlds, despite the high mission cost and long preparation time. The Earth's atmosphere, day-night cycles, atmospheric conditions, or turbulence do not affect the observations. It also provides a longer observation window to continuously observe exoplanetary transits and eclipses. Thus, it is worth considering the possibility of a medium-class high-resolution spectroscopy mission. Such a mission could be dedicated to the comprehensive characterization of \textit{known} nearby rocky exoplanets within the habitable zone of their host stars, recommended by Jean-Loup Bertaux as part of the Voyage 2050\cite{tacconi2021voyage} Long-term planning of the European Space Agency Science Program. One of the challenges lies in the fact that traditional high-resolution spectrographs, which are typically based on Echelle e.g. EarthFinder \cite{Plavchan2020} or immersion gratings, are bulky and primarily suitable for the largest and most costly flagship space missions. As an alternative approach, the Virtually Imaged Phased Array (VIPA)\cite{VIPA}, holds significant promise for various astronomical applications where ultra-high resolution and precise calibration are essential. \edited{Figure \ref{fig:Fig6} illustrates the relationship between total photon efficiency—defined as the product of the telescope's collecting area and spectrograph throughput—and the spectrograph resolution for narrow-band observations to demonstrate the advantages of combining a large telescope aperture, high throughput, and high-resolution instrument to collect photons in hazy and cloudy scenarios more efficiently.} These applications encompass fields such as solar physics research\cite{Zhu_2020}, exoplanet detection using the radial velocity method, and the characterization of exoplanetary atmospheres\cite{Bourdarot_2018}. Multi-channel VIPA spectrographs can be used for the simultaneous observation of multiple biomarkers. An alternative proposal for utilizing interferometry-based instruments in space involves Imaging Fourier Transform Spectrographs (iFTS). These instruments are generally less susceptible to detector noise and offer several other advantages, including simultaneous imaging and spectroscopy, reduced detector size requirements, and adjustable spectral resolution. By comparing analytic and numerical models of Integral Field Spectrographs (IFS) and iFTS, it is determined that iFTS would be more efficient than an IFS if the readout noise of near-IR detectors is halved from the current state-of-the-art level (readout noise above 2-3 electrons/pix/frame) \cite{Zhang2023}. Lastly, for the study of the atmosphere, especially for terrestrial exoplanets in the temperate zone, large mirror apertures are a necessity. The lightweight segmented space telescopes created through additive manufacturing \cite{Atkins2019} could emerge as a pivotal contributor.

\section{Conclusion}\label{conclusion}
In this work, we provided an overview of current and upcoming technical capabilities for characterizing exoplanet atmospheres around rocky planets. Both direct imaging and transit spectroscopy can greatly benefit from high-resolution, high-throughput spectrographs. We then compared current state-of-the-art on-sky spectra that demonstrate the range of technical capabilities. Our focus was particularly on $\mathrm{O_2}$, which complements already detected atmospheric molecules for a comprehensive characterization of star-planet interaction, geological processes, and potential biological activity. We highlight significant instrumentation advancements and challenges in the field of exoplanet atmosphere studies, particularly focusing on rocky exoplanets. Researchers are continually pushing the boundaries of our understanding of distant worlds through the utilization of grating-based and interferometer-based high-resolution spectroscopy, as well as innovative techniques in high-contrast imaging. The detection and characterization of molecules such as water $\mathrm{H_2O}$, methane $\mathrm{CH_4}$, and oxygen $\mathrm{O_2}$ provide crucial insights into the composition to infer potential habitability of exoplanet atmospheres. However, challenges such as clouds, haze resulting in featureless spectra, and instrumental limitations persist, underscoring the need for further technological innovation and observational refinement. Despite these challenges, advancements in technology and instrument development, particularly with the advent of novel instrument suites on Extremely Large Telescopes (ELTs) and next-generation space-based instruments such as Habitable Worlds Observatory (HWO), offer promising prospects for overcoming these obstacles. 

\edited{With detailed exposure time calculations for detecting $\mathrm{O_2}$, at the same exposure time, a spectral resolution of 300,000 achieves higher significance compared to 100,000. For the optimistic case of clear skies, the exposure time is reduced by 10\%, while in the realistic 50\% haze scenario, the required exposure time decreases by 30\%. In the most challenging haze and cloud scenario, increasing the resolution from R=100,000 to R=300,000 enables significant detection with 4 times shorter exposure time (and thus correspondingly fewer transit). Therefore, HWO should consider incorporating a high-resolution instrument, as recommended in the 2018 Exoplanet Science Strategy report \cite{NAP25187}. Although telluric contamination is not a concern in space, a high-resolution compact instrument would be advantageous for biomarker observations.}

As we continue to unravel the mysteries of exoplanetary atmospheres, each discovery brings us closer to unlocking the secrets of other worlds and exploring the potential for life beyond our solar system. Collaboration among astronomers and engineers remains crucial for making significant strides in this field.

\small
\bibliography{ref_HRS}

%\noindent LaTeX formats citations and references automatically using the bibliography records in your .bib file, which you can edit via the project menu. Use the cite command for an inline citation, e.g.  \cite{Hao:gidmaps:2014}.

%For data citations of datasets uploaded to e.g. \emph{figshare}, please use the \verb|howpublished| option in the bib entry to specify the platform and the link, as in the \verb|Hao:gidmaps:2014| example in the sample bibliography file.

\section*{Acknowledgements}
Author acknowledge the use of the \texttt{ExoAtmospheres Database}, Exoplanet Encyclopedia and NASA Exoplanet Archive database during the preparation of this work. The X-Shooter spectra are based on data obtained from the ESO Science Archive Facility with DOI(s): https://doi.eso.org/10.18727/archive/71. S.R. thanks Johannes Buchner for reading the manuscript. 

\section*{Author contributions statement}
S.R. contributed to all aspects of the manuscript.

\section*{Additional information}

\subsection*{Competing interests} 
The author declares no competing interests.
\subsection*{Data and code availability}
The datasets used and/or analyzed are available from the corresponding author on reasonable request. 

%The corresponding author is responsible for submitting a \href{http://www.nature.com/srep/policies/index.html#competing}{competing interests statement} on behalf of all authors of the paper. This statement must be included in the submitted article file.

\end{document}